\documentclass{article}
\usepackage{amsmath}
\usepackage{graphics}
\usepackage{xcolor}

\newcommand{\bea}{\begin{eqnarray*}}
\newcommand{\eea}{\end{eqnarray*}}
\newcommand{\bean}{\begin{eqnarray}}
\newcommand{\eean}{\end{eqnarray}}
\newcommand{\eqs}[1]{Eqs. (\ref{#1})}
\newcommand{\eq}[1]{Eq. (\ref{#1})}
\newcommand{\meq}[1]{(\ref{#1})}
\newcommand{\fig}[1]{Fig. \ref{#1}}
\newcommand{\ppa}[2]{\left(\frac{\partial}{\partial #1}\right)^{#2}}

\newcommand{\eqn}{&=&}
\newcommand{\non}{\nonumber \\}
\newcommand{\oh}{\frac{1}{2}}

\title{Static spherically symmetric thin shell wormhole colliding with a spherical thin shell}
\author{ Xiaobao Wang\thanks{Email: xiaobao@mail.bnu.edu.cn} and Sijie Gao\thanks{Corresponding author. Email: sijie@bnu.edu.cn} \\
Department of Physics, Beijing Normal University,\\
Beijing 100875, China}
\begin{document}
\maketitle

\begin{abstract}
We consider a static spherically symmetric thin shell wormhole that collides with another thin shell consisting of ordinary matter. By employing the geometrical constraint, which leads to the conservation of energy and momentum, we show that the state after the collision can be solved from the initial data. In the low speed approximation, the solutions are rather simple. The shell may either bounce back or pass through the wormhole. In either case, the wormhole shrinks right after the collision. In the ``bouncing'' case, a surprising result is that the radial speeds before and after the collision satisfy an addition law, which is independent of other parameters of the wormhole and the shell.  Once the shell passes through the wormhole, we find that the shell always expands. However, the expansion rate is the same as its collapsing rate right before the collision. Finally, we find out the solution for the shell moving together with the wormhole. This work sheds light on the interaction between wormholes and matter.


\end{abstract}

\section{Introduction}
The thin shell model\cite{israel-66} is an idealization of the real matter distribution and has given many interesting solutions in general relativity and alternative gravity theories. Using the ``cut and paste'' technique, Visser\cite{visser} proposed a simple method to construct thin shell wormholes. Linear stability of thin shell wormholes was studied later \cite{ernesto}-\cite{yue}. It is well known that a wormhole must contain exotic matter, which violates some energy conditions.

Usually, a wormhole is treated as a fixed background.
It is important and interesting to know how a wormhole interacts with matter. For instance, how will the wormhole change when a self-gravitating object falls into it? This is a difficult issue because it involves backreaction.  In this paper, we use a thin shell consisting of ordinary matter as the source of perturbation. We study the physical process when an initially static wormhole collides with the ordinary shell. Investigating this process may help us understand the stability and   traversability of wormholes. Although the thin shell is a simplified object, studying the problem of collision is still not easy.
Notice that Langlois, Maeda and Wands \cite{prl02} derived the conservation laws in the collision of thin shells. This is a consequence of the continuity of the spacetime metric. By employing the LMW mechanism as well as the Israel junction condition, we show that the problem can be solved at low speed limits. We first consider two interesting scenarios of collision:  The shell bounces back or the shell passes through the wormhole. The two scenarios obey different equations of motion. For the ``bouncing'' case, we found that the radial speeds of the wormhole and the shell after the collision are both proportional to the original speed of the shell. Most interestingly, the three speeds satisfy a simple addition law. For the ``passing'' case, our solution is consistent with that in \cite{prd13} which was obtained by a different approach. We also discuss the scenario that the shell and the wormhole stick together after the collision.

\section{General properties for shell collisions} \label{sec-r}
In this section, we review the LMW mechanism, which shows that the continuity of spacetime implies the conservation laws.

\begin{figure}[htmb]
\centering \scalebox{0.7} {\includegraphics{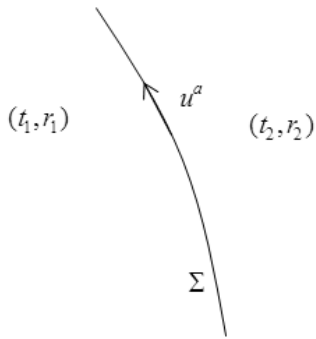}}
\caption{A spherical shell $\Sigma$ moving with four-velocity $u^a$.} \label{fig-tr}
\end{figure}

Consider a spherical shell $\Sigma$ moving in a spherical spacetime. The coordinates on the two sides of the shell are labeled by $(t_1, r_1)$ and $(t_2,r_2)$, where we have dropped the $(\theta, \phi)$ coordinates for simplicity (see \fig{fig-tr}).

The metrics on both sides are of the form
\bean
ds_i^2=-f_i(r)dt^2+f_i^{-1}(r)dr^2+r^2 d\Omega^2
\eean
where $i=1,2$ and in the Schwarzschild case
\bean
f_i(r)=1-\frac{2M_i}{r}
\eean
Let $[K_{ab}]$ be the jump of the extrinsic curvature across the shell. The evolution of the shell follows the junction condition \cite{israel-66}
\bean
[K_{ab}]=-8\pi\left(S_{ab}-\oh S h_{ab}\right)\,,
\eean
where $h_{ab}$ is the induced metric on $\Sigma$ and $S_{ab}$ is the energy-momentum tensor of the shell. In the spherical case, we have
\bean
\epsilon_2\sqrt{f_2+\dot r^2}-\epsilon_1\sqrt{f_1+\dot r^2}=4\pi\sigma r \,,\label{ijun}
\eean
where $\epsilon_i=-1 $ corresponds to $r$ increasing from left to right in \fig{fig-tr} and $\epsilon_i=+1$ otherwise. For an ordinary shell, $\epsilon_1=\epsilon_2=-1$ , while for a wormhole with exotic matter, $\epsilon_1=1$ and $\epsilon_2=-1$.

Note that $r_1=r_2$ by continuity, but $t$ is discontinuous across the shell.  We may write the four-velocity of the shell as
\bean
u^a=\dot t_i \ppa{t_i}{a}+\dot r \ppa{r_i}{a}
\eean
Note that we have used $\dot r$ instead of $\dot r_i$ because $r_1=r_2$. The normalization condition $g_{ab}u^a u^b=-1$ yields
\bean
\dot t_i=\pm \sqrt{\frac{f_i+\dot r^2}{f_i^2}} \label{tdi}
\eean
The normal vector of $\Sigma$ is of the form
\bean
n_i^a=\frac{\dot r}{f_i(r)}\ppa{t_i}{a}+\sqrt{\dot r^2+f_i}\ppa{r_i}{a}
\eean

Now we have three orthogonal and normal tetrads related by the following Lorentz transformation \cite{prl02}
\bean
 \left(\begin{array}{c}
u^a \\
n^a \end{array} \right)=\Lambda(\alpha_{i})\left(\begin{array}{c}
\sqrt{\frac{1}{f_i}}\ppa{t_i}{a} \\
\sqrt{f_i} \ppa{r_i}{a} \end{array} \right)
\eean
where
\bean
\Lambda(\alpha)=\left(\begin{array}{cc}
\cosh(\alpha) & \sinh(\alpha) \\
\sinh(\alpha) & \cosh(\alpha) \end{array} \right)
\eean
and
\bean
\alpha_i=\sinh^{-1}\frac{\epsilon_i\dot r}{\sqrt{f_i}}
\eean
Therefore,
\bean
\left(\begin{array}{c}
\sqrt{\frac{1}{f_2}}\ppa{t_2}{a} \\
\sqrt{f_2} \ppa{r_2}{a} \end{array} \right)=\Lambda(\alpha_1-\alpha_2)\left(\begin{array}{c}
\sqrt{\frac{1}{f_1}}\ppa{t_1}{a} \\
\sqrt{f_1} \ppa{r_1}{a} \end{array} \right) \label{loren}
\eean

\begin{figure}[htmb]
\centering \scalebox{0.7} {\includegraphics{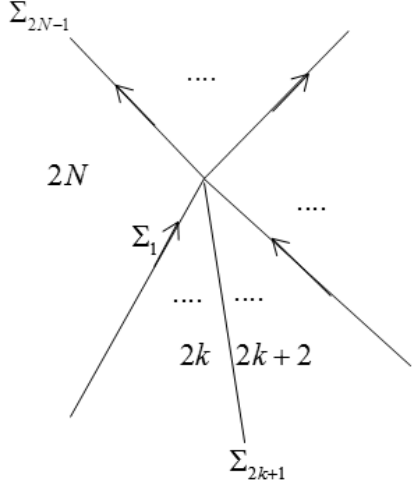}}
\caption{$M$ shells colliding at a moment and $N-M$ shells appearing after the collision.} \label{Nshell}
\end{figure}
Now consider $M$ shells colliding simultaneously. After the collision, $N-M$ shells appear. So there are $N$ shells in total at the spacetime point of collision (see \fig{Nshell}). We label each shell by an odd number and the region in between by an even number.

Define the angles on each side of the shell by
\bean
\sinh\alpha_{2k-1|2k}=\frac{\epsilon_{2k}\dot r_{2k-1}}{\sqrt{f_{2k}}} \label{dal}
\eean

Now perform the Lorentz transformation \meq{loren} to each shell in \fig{Nshell} associated with $\alpha_i$  near the collision point. Then after completing this process, we end up with the consistent relation
\bean
\Pi_{k=1}^{N}\Lambda(\alpha_{2k-1|2k}-\alpha_{2k-1|2k-2})=1
\eean
which is equivalent to
\bean
\sum_{i=1}^{2N}\alpha_{i|i+1}=0 \label{calpha}
\eean
where we have defined $\alpha_{2k|j}\equiv -\alpha_{j|2k}$. This is an important result of \cite{prl02}. It is also called the geometrical constraint, which reflects the continuity of the metric at the collision point. One can show that \eq{calpha}, together with the junction conditions, indicates the conservation of energy and momentum.

\section{Collision of a static wormhole with a shell }

\begin{figure}[htmb]
\centering \scalebox{0.7} {\includegraphics{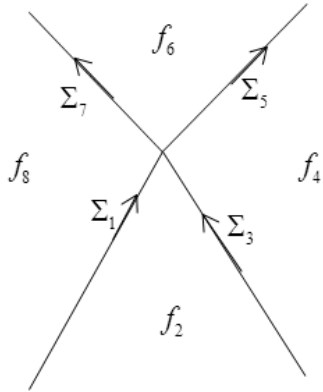}}
\caption{A wormhole $\Sigma_1$ colliding with a shell $\Sigma_3$. The shell remains on the ``right'' side of the wormhole after the collision. So $\Sigma_5$ represents the shell and $\Sigma_7$ represents the wormhole.} \label{fig-4shell}
\end{figure}

Suppose that the spherical thin shell wormhole is originally at rest ($\Sigma_1$ in
\fig{fig-4shell}). Then another thin shell $\Sigma_3$ with ordinary matter collapses and hits the wormhole. After the collision, a new spacetime region with function $f_6$ in \fig{fig-4shell} emerges. This is the case that $M=2$ and $N=4$ in \fig{Nshell}.
 Given initial values, we  show that the junction conditions and the consistency condition \meq{calpha} are just enough to determine the state after the collision. We are interested in two scenarios after the collision: The shell either remains on the same side of the wormhole or passes through the wormhole. It is reasonable to assume that the proper masses of the wormhole and the shell remain unchanged just after the collision. The two cases are determined by different sets of equations, so we shall solve them one by one. Another scenario is that the shell and the wormhole move together after the collision. In this case, the speed and the mass of the new wormhole can both be solved.

\subsection{Case 1: The bouncing solution}
In this case, the shell remains on the same side of the wormhole after the collision, as depicted in \fig{fig-4shell}.
According to \eq{dal}, the Lorentz angles are given by
\bean
\sinh\alpha_{18}\eqn 0, \ \ \ \sinh\alpha_{12}= 0 \label{sinh1} \\
\sinh\alpha_{32}\eqn -\frac{\dot r_3}{\sqrt{f_2}}, \ \ \ \sinh\alpha_{34}= -\frac{\dot r_3}{\sqrt{f_4}}\label{sinh2} \\
\sinh\alpha_{54}\eqn -\frac{\dot r_5}{\sqrt{f_4}}, \ \ \ \sinh\alpha_{56}= -\frac{\dot r_5}{\sqrt{f_6}} \label{sinh3} \\
\sinh\alpha_{76}\eqn -\frac{\dot r_7}{\sqrt{f_6}}, \ \ \ \sinh\alpha_{78}= \frac{\dot r_7}{\sqrt{f_8}} \label{sinh4}
\eean
The consistency condition is given by
\bean
\alpha_{18}-\alpha_{12}+\alpha_{32}-\alpha_{34}+\alpha_{54}-\alpha_{56}+\alpha_{76}
-\alpha_{78}=0 \label{azero}
\eean

It is reasonable to assume that the masses of the wormhole and the shell remain unchanged, i.e.,
\bean
\rho_7=\rho_1,~~~~ \rho_5=\rho_3
\eean
For simplicity, we assume that the static wormhole is symmetric. Hence,
\bean
\dot r_1\eqn 0 \\
f_2\eqn f_8
\eean

Applying \eq{ijun} to each shell, we have the following four equations:
\bean
-2\sqrt{f_2}\eqn \tilde\rho_1  \label{f1}\\
\sqrt{f_2+\dot r_3^2}-\sqrt{f_4+\dot r_3^2}\eqn \tilde\rho_3  \label{f2} \\
\sqrt{f_6+\dot r_5^2}-\sqrt{f_4+\dot r_5^2}\eqn \tilde\rho_3 \label{f3} \\
-\sqrt{f_2+\dot r_7^2}-\sqrt{f_6+\dot r_7^2}\eqn \tilde\rho_1  \label{f4}
\eean

We may choose the initial values  $f_4(<f_2)$ and $\tilde\rho_3$. Then $\rho_1$ and $\dot r_3$ can be solved from \eqs{f1} and \meq{f2}. By \eqs{f3}, \meq{f4}, and \meq{azero}, one can solve for $\dot r_5$, $\dot r_7$, and $f_6$.

These equations are not easy to solve, even numerically. However, in the following subsections, we derive an inequality for $f_6$ and then find the solutions at low speed limits.

\subsubsection{An inequality}
Combination of \eq{f1} and \eq{f4} yields
\bean
2\sqrt{f_2}=\sqrt{f_2+\dot r_7^2}+\sqrt{f_6+\dot r_7^2}
\eean
Solving for $\dot r_7^2$, we have
\bean
\dot r_7^2\eqn \frac{1}{16 f_2}(f_6^2-10 f_2 f_6+9f_2^2) \non
\eqn \frac{f_2}{16}(p^2-10p+9)
\eean
where $p=\frac{f_6}{f_2}$.
Since $\dot r_7^2>0$ , we must have

\bean
f_6< f_2 ~~~ \textrm{or} ~~~ f_6>9f_2
\eean
This is the range for $f_6$. In the following, we consider a solution of perturbation, where $f_6$ is very close to $f_2$. Thus, only $f_6<f_2$ will be considered.

\subsubsection{Solutions for low speeds}
When treating the behavior of the shell as a perturbation, it is reasonable to assume that $\dot r_3$ is small. Consequently, $\dot r_5$ and $\dot r_7$ are also small.
 Then by Taylor expansion, \eqs{f1}--\meq{f4} and \eq{azero} may be approximated as
\bean
-2\sqrt{f_2}\eqn \tilde\rho_1  \label{rr1} \\
\sqrt{f_2}+\frac{\dot r_3^2}{2\sqrt{f_2}}-\sqrt{f_4}-\frac{\dot r_3^2}{2\sqrt{f_4}} \eqn\tilde\rho_3  \label{rr2}\\
\sqrt{f_6}+\frac{\dot r_5^2}{2\sqrt{f_6}}-\sqrt{f_4}-\frac{\dot r_5^2}{2\sqrt{f_4}} \eqn\tilde\rho_3    \label{rr3}\\
-\sqrt{f_2}-\frac{\dot r_7^2}{2\sqrt{f_2}}-\sqrt{f_6}-\frac{\dot r_7^2}{2\sqrt{f_6}} \eqn\tilde\rho_1  \label{rr4}
\eean
and
\bean
-\frac{\dot r_3}{\sqrt{f_2}}+\frac{\dot r_3}{\sqrt{f_4}}-\frac{\dot r_5}{\sqrt{f_4}}+\frac{\dot r_5}{\sqrt{f_6}}-\frac{\dot r_7}{\sqrt{f_6}}-\frac{\dot r_7}{\sqrt{f_2}}=0 \label{a6}
\eean

These equations are still not straightforward to solve.
First notice that \eqs{rr1} and \meq{rr4} yield
\bean
\sqrt{f_2}-\sqrt{f_6}=\frac{\dot r_7^2}{2\sqrt{f_2}}+\frac{\dot r_7^2}{2\sqrt{f_6}}\equiv x>0 \,. \label{xr}
\eean
So
\bean
\sqrt{f_6}=\sqrt{f_2}-x
\eean
Now, $\dot r_5^2$ and $\dot r_7^2$ and $x$ are in the same order. So we can replace $\sqrt{f_6}$ in the denominator of \eqs{rr3} by $\sqrt{f_2}$  and obtain
\bean
 \sqrt{f_2}-x+\frac{\dot r_5^2}{2\sqrt{f_2}}-\sqrt{f_4}-\frac{\dot r_5^2}{2\sqrt{f_4}} \eqn \sqrt{f_2}+\frac{\dot r_3^2}{2\sqrt{f_2}}-\sqrt{f_4}-\frac{\dot r_3^2}{2\sqrt{f_4}}
 \eean
i.e.,
\bean
 -x+\frac{\dot r_5^2}{2\sqrt{f_2}}-\frac{\dot r_5^2}{2\sqrt{f_4}} \eqn +\frac{\dot r_3^2}{2\sqrt{f_2}}-\frac{\dot r_3^2}{2\sqrt{f_4}} \label{mx}
\eean
Using the same approximation, \eq{xr} becomes
\bean
x=\frac{\dot r_7^2}{\sqrt{f_2}}
\eean
Substitution into \eq{mx} yields
\bean
 -\frac{\dot r_7^2}{\sqrt{f_2}}+\frac{\dot r_5^2}{2\sqrt{f_2}}-\frac{\dot r_5^2}{2\sqrt{f_4}} \eqn +\frac{\dot r_3^2}{2\sqrt{f_2}}-\frac{\dot r_3^2}{2\sqrt{f_4}}  \label{ww1}
\eean
Similarly, \eq{a6} becomes
\bean
 -\frac{\dot r_3}{\sqrt{f_2}}+\frac{\dot r_3}{\sqrt{f_4}}-\frac{\dot r_5}{\sqrt{f_4}}+\frac{\dot r_5}{\sqrt{f_2}}-2\frac{\dot r_7}{\sqrt{f_2}}\eqn 0 \label{ww2}
\eean
Let
\bean
k=\frac{\sqrt{f_4}}{\sqrt{f_2}}<1
\eean
 Then the solution of \eqs{ww1} and \meq{ww2} is given by
\bean
\dot r_5\eqn \frac{1-3k}{k+1}\dot r_3  \label{sr5} \\
\dot r_7\eqn \frac{2-2k}{k+1}\dot r_3 \label{sr7}
\eean
 Since
\bean
-1<\frac{1-3k}{k+1}<1
\eean
for $0<k<1$, we find
\bean
 |\dot r_5|<|\dot r_3|.
\eean
Because $\dot r_3<0$, \eq{sr7} shows
\bean
\dot r_7<0 \,.
\eean
Therefore, the throat of the wormhole always decreases right after the collision.

\eq{sr5} shows that $\dot r_5<0$ when $k<\frac{1}{3}$, which means both the shell and the wormhole shrink. It is easy to find
\bean
\dot r_5-\dot r_7=-\dot r_3>0
\eean
which means
\bean
|\dot r_7|>|\dot r_5|
\eean
if $\dot r_5<0$. So the wormhole shrinks faster than the shell. This is an expected result. Otherwise, the shell cannot remain on the same side of the wormhole after the collision, as shown in \fig{fig-4shell}.

The relation
\bean
\dot r_3+\dot r_5=\dot r_7  \label{r35}
\eean
is surprisingly simple, which means that the sum of the speeds of the shell before and after the collision is equal to the speed of the wormhole after the collision!

The above analysis can be verified by numerical calculation. We choose
\bean
f_2=0.9, ~~~f_4=0.7,~~~\dot r_3=-10^{-5}.
\eean
Then \eqs{rr1}--\meq{a6} can be solved numerically:
\bean
\dot r_5=8.745\times 10^{-6},~~~ \dot r_7=-1.255\times 10^{-6}, ~~~ \sqrt{f_6}=\sqrt{f_2}-1.660\times 10^{-12}\,.
\eean
These results yield
\bean
\frac{\dot r_3+\dot r_5}{\dot r_7}=1.00005\times 10^{-12}
\eean
Hence, \eq{r35} is confirmed.

\subsubsection{Another solution?}
There is another obvious solution:
\bean
\dot r_5=\dot r_3,~~~ \dot r_7=0, ~~~ f_6=f_2\,. \label{tri}
\eean
The apparent interpretation of this solution is that the wormhole remains static and the shell is still collapsing with the same speed. This set of solutions even satisfies the original equations without any approximation. However, this solution is not real in physics. If the wormhole remains static and the shell remains on the same side, the shell must bounce back with a larger radius because the throat of the wormhole has the minimum radius. But this results in $\dot r_5>0$, disagreeing with $\dot r_5=\dot r_3$.  One may think that this solution indicates that the shell passes through the wormhole. If this is the case, the equations must be modified (see the next subsection). In the new configuration, we  see that \eq{tri} is no longer a solution. Therefore, in either case, \eq{tri} should be discarded.

\subsection{Case II: Passing through the wormhole}
Now we assume that the shell travels through the wormhole and appears on the other side. In this case, the positions of $\Sigma_7$ and $\Sigma_5$ should have exchanged their roles in \fig{fig-4shell}. For the ``passing through'' solution, $\Sigma_5$ in \fig{fig-4shell} represents the wormhole after collision and  $\Sigma_7$ represents the shell.

Then \eqs{f1}-\meq{f4} are modified as
\bean
-2\sqrt{f_2}\eqn \tilde \rho_1  \label{fq1}\\
\sqrt{f_2+\dot r_3^2}-\sqrt{f_4+\dot r_3^2}\eqn \tilde\rho_3  \label{fq2} \\
-\sqrt{f_6+\dot r_5^2}-\sqrt{f_4+\dot r_5^2}\eqn \tilde\rho_1  \label{fq3} \\
\sqrt{f_6+\dot r_7^2}-\sqrt{f_2+\dot r_7^2}\eqn \tilde\rho_3  \label{fq4}
\eean
\eqs{sinh1}-\meq{sinh4} become
\bean
\sinh\alpha_{18}\eqn 0, \ \ \ \sinh\alpha_{12}= 0 \label{si1} \\
\sinh\alpha_{32}\eqn -\frac{\dot r_3}{\sqrt{f_2}}, \ \ \ \sinh\alpha_{34}= -\frac{\dot r_3}{\sqrt{f_4}}\label{si2} \\
\sinh\alpha_{54}\eqn -\frac{\dot r_5}{\sqrt{f_4}}, \ \ \ \sinh\alpha_{56}= \frac{\dot r_5}{\sqrt{f_6}} \label{si3} \\
\sinh\alpha_{76}\eqn \frac{\dot r_7}{\sqrt{f_6}}, \ \ \ \sinh\alpha_{78}= \frac{\dot r_7}{\sqrt{f_2}} \label{si4}
\eean

For small $\dot r_i$, we have
\bean
-2\sqrt{f_2}\eqn \tilde\rho_1 \label{qq1}\\
\sqrt{f_2}-\sqrt{f_4}+\frac{\dot r_3^2}{2}\left( \frac{1}{\sqrt{f_2}} -\frac{1}{\sqrt{f_4}} \right)  \eqn \tilde\rho_3  \label{qq2} \\
-\sqrt{f_6}-\sqrt{f_4}-\frac{\dot r_5^2}{2}\left(\frac{1}{\sqrt{f_4}}+\frac{1}{\sqrt{f_6}} \right)\eqn \tilde\rho_1  \label{qq3} \\
\sqrt{f_6}-\sqrt{f_2}+\frac{\dot r_7^2}{2}\left(\frac{1}{\sqrt{f_6}}-\frac{1}{\sqrt{f_2}} \right)\eqn \tilde\rho_3 \label{qq4}
\eean
and
\bean
-\frac{\dot r_3}{\sqrt{f_2}}+\frac{\dot r_3}{\sqrt{f_4}}-\frac{\dot r_5}{\sqrt{f_4}}-\frac{\dot r_5}{\sqrt{f_6}}+\frac{\dot r_7}{\sqrt{f_6}}-\frac{\dot r_7}{\sqrt{f_2}}=0 \label{sr77}
\eean
From \eqs{qq1} and \meq{qq3},  we have
\bean
2\sqrt{f_2}-\sqrt{f_4}-\frac{\dot r_5^2}{2}\left(\frac{1}{\sqrt{f_4}}+\frac{1}{\sqrt{f_6}} \right)\eqn \sqrt{f_6} \label{fsix}
\eean
Therefore, $\sqrt{f_6}-2\sqrt{f_2}+\sqrt{f_4}$ is a small quantity too.  So we may replace  $\sqrt{f_6}$ in the denominator of \eq{fsix} with $2\sqrt{f_2}-\sqrt{f_4}$ and obtain
\bean
\sqrt{f_6}=2\sqrt{f_2}-\sqrt{f_4}-\frac{\dot r_5^2}{2}\left(\frac{1}{\sqrt{f_4}}+\frac{1}{2\sqrt{f_2}-\sqrt{f_4}} \right)  \label{f66}
\eean
Substituting \eq{f66} into \eqs{qq4} and \meq{sr77} (still replacing $\sqrt{f_6}$ in the denominator with $2\sqrt{f_2}-\sqrt{f_4}$ ), we find two quadratic equations of $\dot r_5$ and $\dot r_7$. Two sets of solutions can be obtained straightforwardly:
\bean
\dot r_5\eqn \frac{(k-1)^2\dot r_3}{1+k}\\
\dot r_7\eqn \frac{(3-k)\dot r_3}{1+k}
\eean
or
\bean
\dot r_5\eqn (1-k)\dot r_3 \label{ss5}  \\
\dot r_7\eqn -\dot r_3  \label{ss7}
\eean
where
\bean
k=\frac{\sqrt{f_4}}{\sqrt{f_2}}<1
\eean
We see $\dot r_5<0$ for both solutions, meaning the wormhole keeps shrinking as expected. In the first solution, $\dot r_7<0$ and $|\dot r_7|<|\dot r_5|$, meaning the wormhole shrinks faster than the shell, which is inconsistent with the passing through picture. Therefore, this solution should be discarded.  In the second solution, $\dot r_7>0$, meaning the shell expands after the collision. This is reasonable because when the shell appears on the other side of the wormhole, its radius must be larger than the radius of the throat of the wormhole.

It is worth mentioning that the same issue has also been discussed in \cite{prd13}. By assuming that the four velocities remain unchanged, i.e.,
\bean
u_5^a=u_1^a, ~~~~ u_7^a=u_3^a  \label{u57}
\eean
the authors obtained
\bean
\dot r_5\eqn \tilde\rho_3\frac{\dot r_3}{\sqrt{f_2}}  \label{nr5} \\
\dot r_7\eqn -\dot r_3 \label{nr7}
\eean
We see that \eq{nr7} is exactly our solution \meq{ss7}, while \eq{nr5} reduces to \eq{ss5} at the low speed limit (see \eq{fq2}). This is not a coincidence because \eq{u57} guarantees the conservation law
\bean
m_1 u_1^a+m_3u_3^a=m_1u_5+m_3 u_7^a
\eean
which, as we have mentioned, can be derived from the consistency condition \meq{azero}. However, \eq{u57} only works for the ``passing'' case, not the ``bouncing'' case.

\subsection{Case III: Moving together}
\begin{figure}[htmb]
\centering \scalebox{0.7} {\includegraphics{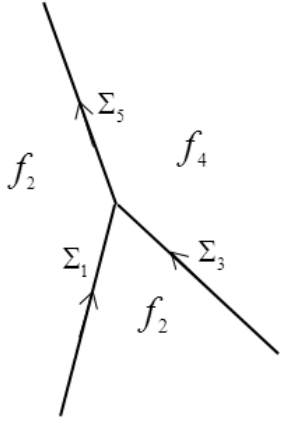}}
\caption{A new wormhole $\Sigma_5$ forms after the collision. } \label{shell3}
\end{figure}

Finally, we consider $M=2$ and $N=3$ in \fig{Nshell}.  This is the case where the shell and the wormhole stick together after the collision (see \fig{shell3} ). We  solve for the radial speed $\dot r_5$ as well as the density $\tilde\rho_5$.

 The junction conditions for the three shells are
\begin{equation}
\tilde\rho_1 =-2\sqrt{f_2}  \label{qw1}
\end{equation}
According to \fig{shell3}
\begin{equation}
\sqrt{f_2+\dot r_3^2}-\sqrt{f_4+\dot r_3^2}=\tilde \rho_3  \label{qw2}
\end{equation}
\begin{equation}
-\sqrt{f_2+\dot r_5^2}-\sqrt{f_4+\dot r_5^2}=\tilde \rho_5   \label{qw3}
\end{equation}

The Lorentz angles are given by
\begin{eqnarray}
\alpha_{32}&=&-\frac{\dot r_3}{\sqrt{f_2}}\\
\alpha_{34}&=&-\frac{\dot r_3}{\sqrt{f_4}}\\
\alpha_{54}&=&-\frac{\dot r_5}{\sqrt{f_4}}\\
\alpha_{56}&=&\frac{\dot r_5}{\sqrt{f_2}}
\end{eqnarray}
Then in the low speed approximation, the consistency condition reads
\begin{equation}
\frac{\dot r_3}{\sqrt{f_2}}-\frac{\dot r_3}{\sqrt{f_4}}+\frac{\dot r_5}{\sqrt{f_4}}
+\frac{\dot r_5}{\sqrt{f_2}}=0  \label{qw4}
\end{equation}
We can easily find
\begin{equation}
\dot r_5=\frac{1-k}{k+1}\dot r_3
\end{equation}
So $\dot r_5<0$, meaning that the new wormhole still shrinks.
It is also easy to obtain
\bean
\tilde\rho_5\eqn -\sqrt{f_2+(\frac{k-1}{k+1}\dot r_3)^2}-\sqrt{f_4+(\frac{k-1}{k+1}\dot r_3)^2} \\
&\approx&-\sqrt{f_2}-\sqrt{f_4}-\frac{s^2\dot r_3^2}{2\sqrt{f_2}}-\frac{s^2\dot r_3^2}{2\sqrt{f_4}}
\eean
where $s=\frac{1-k}{1+k}$. So we can obtain
\bean
\tilde\rho_5=\tilde\rho_1+\tilde\rho_3-\frac{(1+s^2)\dot r_3^2}{2\sqrt{f_2}}-\frac{(1+s^2)\dot r_3^2}{2\sqrt{f_4}}
\eean
This is the relation for the densities, which shows
\bean
\tilde\rho_5<\tilde\rho_1+\tilde\rho_3
\eean
or
\bean
|\tilde\rho_5|>|\tilde\rho_1|-\tilde\rho_3\,.
\eean
The last inequality is due to the fact that $\tilde\rho_1<0$ and $\tilde\rho_5<0$.

\section{Conclusions}
We have investigated the collision of a static wormhole with a spherical thin shell containing ordinary matter. The junction condition of thin shells and the geometrical constraint at the collision event play crucial roles in the process. When the shell hits the wormhole at a low speed, the radial speeds of the wormhole and the shell after the collision are proportional to the initial speed. When the shell bounces back, we have found that the radial speed of the wormhole after collision is equal to the sum of the speeds of the shell before and after the collision. This result has been verified numerically. If the shell goes through the wormhole, our results are consistent with those in \cite{prd13}. Our methods apply to a wide class of spherical spacetimes and can be generalized to other collisions, for instance, the collision of two moving thin shells. We have been focusing on calculating the data right after the collision. The evolutions of the shells after the collision depend on the equations of states. Our calculation provides the initial data for the evolutions.

\section*{Acknowledgements}
 This research was supported by NSFC Grants No. 11235003, No. 11375026 and NCET-12-0054.

\end{document}